\documentclass[fleqn,twoside,twocolumn,nofootinbib]{revtex4} % Specifies the document class %,unsortedaddress
\usepackage{ujp} % \usepackage[cyr]{ujp} for cyrillic, \usepackage[web]{ujp} for web
\begin{document}
\title[MACROPARTICLE MOVEMENT VELOCITY]%колонтитул
{MACROPARTICLE MOVEMENT VELOCITY IN DUSTY STRUCTURES OF VARIOUS COMPOSITIONS}%
\author{A.D. KHAKHAEV}%
\affiliation{Petrozavodsk State University, Research and Educational Center on Basic Problems \\
of Application of Low Temperature Plasma Physics }%
\address{10, Universitetskaya Str., Petrozavodsk 185910, Russia}%
\email{piskunov@plasma.karelia.ru}%e-mail
\author{A.A. PISKUNOV}%
\affiliation{Petrozavodsk State University, Research and Educational Center on Basic Problems \\
of Application of Low Temperature Plasma Physics }%
\address{10, Universitetskaya Str., Petrozavodsk 185910, Russia}%
\email{piskunov@plasma.karelia.ru}%e-mail
\author{S.F. PODRYADCHIKOV}%
\affiliation{Petrozavodsk State University, Research and Educational Center on Basic Problems \\
of Application of Low Temperature Plasma Physics }%
\address{10, Universitetskaya Str., Petrozavodsk 185910, Russia}%
\email{piskunov@plasma.karelia.ru}%e-mail
\udk{???}  \razd{\secvii}

\setcounter{page}{1284}%
\maketitle

\begin{abstract}
The results of experimental investigations of the movement velocity
of a macroparticle in the dusty structures of various
physical-chemical compositions formed in a stratified column of a dc
glow discharge, are presented. The macroparticle substances are
alumina ($r=10-35~\mu$m), polydisperse Zn ($r=1-20~\mu$m) and Zn$'$
($r=20-35~\mu$m). Plasma-forming gases are inert gases (Ne, Ar). The
inverse relation between the velocity and the gas pressure (in the
range 40--400 Pa) is found and, for the same material of
macroparticles in different gas plasmas, is confirmed by theory and
does not contradict observations. But, to explain a difference of
quantitative data for macroparticles made from different materials
in Ar plasma, the additional research is required.
\end{abstract}

\vskip1cm \noindent Complex plasma is a widespread state of matter
in the nature and technological operations. The problem of
particles' kinetics in complex plasma arises, because the appearance
of macroparticles in plasma modifies plasma properties and gives
rise to various processes: an interaction between macroparticles and
ions and electrons gives rise to the charging of macroparticles,
power from plasma to macroparticles transfers through plasma
oscillations, macroparticles lose energy because of neutral impacts,
negative charge of macroparticles gives rise to forming the particle
flows in plasma, {\it etc}. The investigation of these and many
other processes has a high significance for power machines
(tokamaks, lasers, rocket engines, {\it etc}.), plasma technologies
of material processing and producing, astrophysics, for the physical
modeling of condensed matter, microcosmic objects, and astrophysical
objects inaccessible for experiments.\looseness=1

The plasma-dust structures in complex plasma represent an unusual
state of matter. We may expect that a variation of the
physical-chemical characteristics of a plasma-forming gas and a
macroparticle material will result in quantitative changes of the
characteristics of a structure, as well as in a macroparticle motion
\cite{Bu, Kha} under similar conditions.

It is known from the previous experiments that the macroparticles
embedded into plasma acquire a kinetic temperature much more than
the temperature of ambient gas, i.e., the interaction between
charged macroparticles and ambient plasma in ordered structures
results in the essential heating of macroparticles \cite{Zh}. In
work \cite{Vau}, the theoretical justification of the heating of
macroparticles due to a fluctuating electric field of plasma is
proposed. An important result of that work is the expression for the
kinetic temperature of macroparticles which is propotional to the
electron temperature $T_e$. However, no quantitative experimental
data, which would confirm it for the complex plasma of a dc glow
discharge, are available.

In present work, we study the influence of the physical-chemical
composition and some parameters of complex plasma on the velocity of
macroparticles. In experiments, we use the complex plasma of a dc
glow discharge at low currents and pressures. The ordered structures
(Fig. 1) are formed from macroparticles levitating in strata of a
positive column.

%Fig 1
\begin{figure}
\includegraphics[width=7cm]{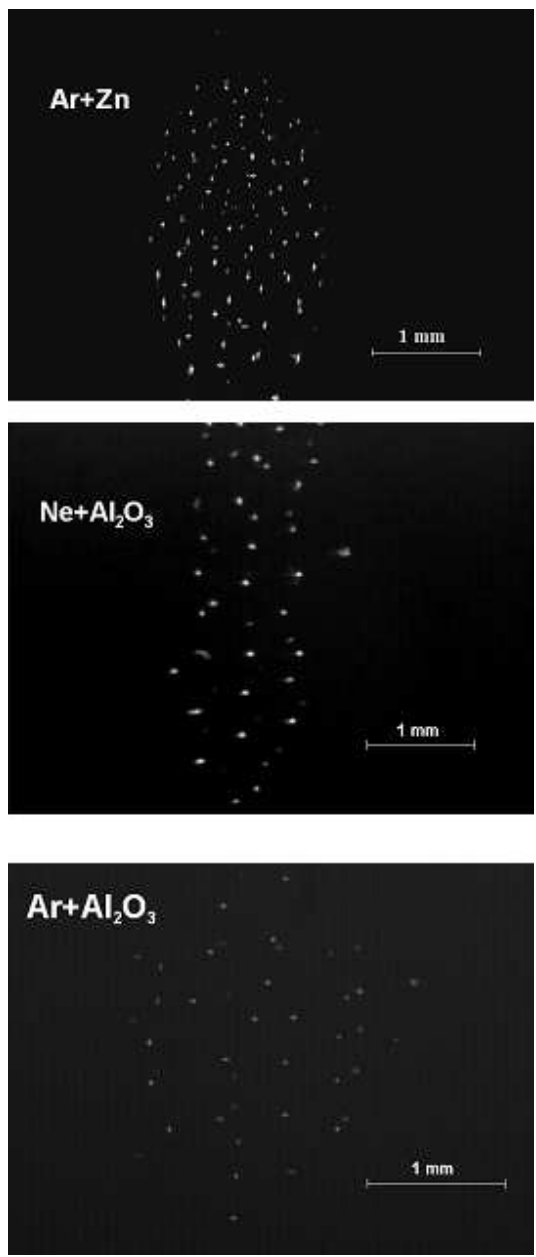}
\vskip-3mm\caption{\label{fig:epsart} Structures of macroparticles in a
stratum glow discharge (a gas pressure of 0.6 Torr, a discharge current of 1
mA)}
\end{figure}

A vertically oriented gas discharge tube (the internal diameter
$2R=30$ mm) is employed \cite{Bu, An}. The shape of electrodes is a
cylinder with a diameter of 25 mm. A narrowing (diameter~2.5 mm) is
placed within the tube. The narrowing is necessary to the generation
of stable standing striations in a wide range of parameters.
Macroparticles are injected in plasma from a container which is
placed at the top of the tube. The macroparticle materials are
Al$_2$O$_3$ ($r=10$--35~$\mu$m, $\langle r\rangle=23~\mu$m),
polydisperse Zn ($r=1$--20~$\mu$m, $\langle r\rangle=8$~$\mu$m), and
Zn' ($r=20$--35~$\mu$m, $\langle r\rangle=28~\mu$m). Plasma-forming
gases are inert gases (Ne, Ar).

Powder is injected into plasma by shaking a container.
Macroparticles fall through the grid bottom and are trapped in
strata. The glass tube is evacuated by a diffusion pump, and the
base vacuum is about $10^{-5}$--$10^{-4}$ Torr. The pressure of a
plasma-forming gas $p$ is varied in the interval 0.3--3 Torr, the
discharge current $I$ is 0.4--1.2 mA (the electron density $n_e$ in
dust-free plasma is about $10^{8}$ cm$^{-3}$, the electron
temperature is about several eV). The probe measurements under
similar conditions can be found in \cite{Khr}.

Macroparticles in a cross-section of the macroparticle structure were
observed by a CCD-camera for 4 s (a picture frequency of 25
Hz). 100 displacements was done by every macroparticle for this
time.  Every displacement $ds_i$ occurs for 1/25 s at a velocity
$v_i$. The macroparticle mean velocity of a displacement for 4
s and the dispersion were calculated for 100 $v_i$.
Then the ensemble averaging for $N$ macroparticles (visible in a
cross-section) was made taking into account the dispersion of
velocity values. Errors were estimated, by basing on 3 measurements for
each point in Fig.~2 and Fig. 3.

%Fig 2
\begin{figure}
\includegraphics[width=\column]{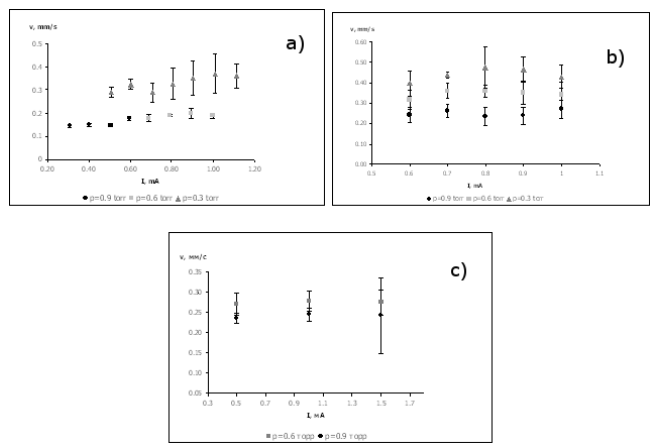}
\vskip-3mm\caption{\label{fig:epsart} Experimental dependence of
the average velocity of motion of macroparticles on the discharge current: {\it
a})Ar+Al$_2$O$_3$ \cite{Kha}, {\it b}) Ne+Al$_2$O$_3$ \cite{Kha},
{\it c}) Ar+Zn}%\vskip3mm
\end{figure}

%Fig 3
\begin{figure}
\includegraphics[width=\column]{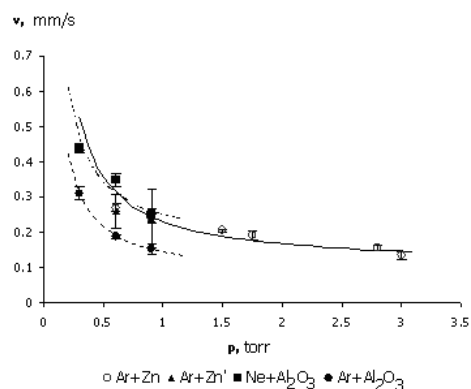}
\vskip-5mm\caption{\label{fig:epsart} Macroparticle velocity
versus the gas pressure (figures -- experiment, lines --
approximation by the inverse relation) at room temperature}
\end{figure}

The slight influence of the electron density on the
macroparticle velocity is evident from the dependence of the mean
macroparticle velocity on the discharge current (Fig.~2).
The discharge current $I \approx en_{e} v_{e}R^{2}$, where
$v_{e}$ is the electron drift velocity.

An inverse relation between the macroparticle velocity and the gas
pressure (Fig. 3) was found for ordered dusty structures of
various compositions. A decrease of the velocity (Fig. 3) is caused
by both the friction on neutrals and a decrease of the electron
temperature. The trend to curves $v(p)$ for similar macroparticles
(Al$_2$O$_3$) in Ar and Ne plasma correlates with curves $T_e(p)$
for the same gas plasma \cite{Gra}. These experimental data are in
agreement with the theory \cite{Vau}. The influence of the electron
temperature on the macroparticle motion was observed, for example,
in experiments with a high-energy electron beam (tens of eV) under
similar discharge conditions \cite{Vas}, but without quantitative
measurements.

The conducted researches give us an experimental evidence that the
macroparticle material also influences, in particular, the kinetic
temperature of macroparticles in ordered dusty plasma structures.
This result is not follow from \cite{Vau} and cannot be also
explained by the friction on neutrals, because the macroparticles
have approximately identical size (Al$_2$O$_3$ and Zn$'$). To
explain a difference of quantitative data for different materials of
macroparticles in Ar plasma, the additional research is required.

\vskip3mm This work was partially supported by CRDF on Grants
RUX0-013-PZ-06/BF7M13, the Ministry of Education and Science of the
Russian Federation; the Government of Karelia.

\rezume{%
ШВИДКІСТЬ ПЕРЕМІЩЕННЯ МАКРОЧАСТИНОК\\ У ПИЛОВИХ СТРУКТУРАХ РІЗНОГО\\
КОМПОНЕНТНОГО СКЛАДУ}{А.Д. Хахаєв, А.А. Піскунов, С.Ф. Подрядчиков}
{Наведено результати експериментальних досліджень швидкості
переміщення мікрочастинок у пилових структурах різного
фізико-хімічного складу. Плазмово-пилові структури формувались у
стратах позитивного стовпа жевріючого розряду. У ролі пилової
компоненти використовували частинки оксиду алюмінію ($r=10$--35 мкм)
і цинку ($r=1$--20 мкм і $r=20$--35 мкм). Газовими наповнювачами
виступали інертні гази (Ne, Ar). Було отримано зворотні залежності
швидкості переміщення мікрочастинок від тиску газу (в діапазоні
40--400 Па). При цьому для мікрочастинок із однакового матеріалу,
але у плазмі різних газів дані узгоджуються з теоретичними
передбаченнями і не суперечать спостереженням. Але, щоб пояснити
розбіжність у швидкостях макрочастинок різного матеріалу в аргоновій
плазмі, потрібні додаткові дослідження.}

\end{document}